%% file: main.tex
\pdfoutput=1
\documentclass[12pt,a4paper]{article}
\usepackage{ifthen} % for conditional statements
\newboolean{pdflatex}
\setboolean{pdflatex}{true} % False for eps figures 
\usepackage{relsize}

\newboolean{articletitles}
\setboolean{articletitles}{true} % False removes titles in references

\newboolean{uprightparticles}
\setboolean{uprightparticles}{false} %True for upright particle symbols

\def\paperauthors{LHCb collaboration} % Leave as is for PAPER, CONF and FIGURE
\def\paperasciititle{Polarization measurement of Lambdac baryons in pNe collisions at sqrt(s)NN=68.6 GeV} % Set ASCII title here !! MAKE sure it's only ASCII characters !! 
\def\papertitle{Polarization measurement of %$\overset{\text{\relsize{-1}(}-\text{\relsize{-1})}}
$\PLambda_c^+$ and $\offsetoverline{\PLambda}{}^-_c$ baryons in $p$Ne collisions at \mbox{$\ensuremath{\protect\sqrt{s_{\scriptscriptstyle\text{NN}}}} = 68.6 \aunit{Ge\kern -0.1em V}$} } % Latex formatted title
\def\paperkeywords{{High Energy Physics}, {LHCb}} % Comma separated list
\def\papercopyright{\the\year\ CERN for the benefit of the LHCb collaboration} % new since 9/Apr/2018
\def\paperlicence{CC BY 4.0 licence}
\def\paperlicenceurl{https://creativecommons.org/licenses/by/4.0/}

\newif\ifEnableSectionTOCLinks
\EnableSectionTOCLinksfalse % deactivated

\input{preamble}
\usepackage{longtable} % only for template; not usually to be used in PAPERs
\usepackage{booktabs}

\input{thesis-symbols}

\begin{document}

%%%%%%%%%%%%%%%%%%%%%%%%%
%%%%% Title     %%%%%%%%%
%%%%%%%%%%%%%%%%%%%%%%%%%
\renewcommand{\thefootnote}{\fnsymbol{footnote}}
\setcounter{footnote}{1}

% %%%%%%% CHOOSE TITLE PAGE--------
%\onecolumn

\input{title-LHCb-PAPER}

%\twocolumn
% %%%%%%%%%%%%% ---------

\renewcommand{\thefootnote}{\arabic{footnote}}
\setcounter{footnote}{0}

%%%%%%%%%%%%%%%%%%%%%%%%%%%%%%%%
%%%%%  Table of Content   %%%%%%
%%%%%%%%%%%%%%%%%%%%%%%%%%%%%%%%
%%%% Uncomment if desired
%\tableofcontents

\cleardoublepage

%%%%%%%%%%%%%%%%%%%%%%%%%
%%%%% Main text %%%%%%%%%
%%%%%%%%%%%%%%%%%%%%%%%%%

\pagestyle{plain} % restore page numbers for the main text
\setcounter{page}{1}
\pagenumbering{arabic}

%% Uncomment during review phase. 
%% Comment before a final submission.
%\linenumbers

%% This is the main body
%% It is useful to have a single file so comments are not missed in overleaf.
\input{body}

% Do not include this in any draft (just for information in the template)
%\input{acknowledgements_intro}
% Comment this in for paper drafts; do not include this in analysis note, conference and figure reports
\input{acknowledgements}

\input{supplementary-app}

\addcontentsline{toc}{section}{References}
%\setboolean{inbibliography}{true}
\bibliographystyle{LHCb}
\bibliography{main,standard,LHCb-PAPER,LHCb-CONF,LHCb-DP,LHCb-TDR}

\newpage
\input{Authorship_LHCb-PAPER-2025-060}
%\input{~/Downloads/Authorship_LHCb-SD-2024-001-grouped-internal-use-only-do-not-publish-2.tex}

%The author list for journal publications is generated from the
%Membership Database shortly after 'approval to go to paper' has been
%given.  It is available at \url{https://lbfence.cern.ch/membership/authorship}
%and will be sent to you by email shortly after a paper number
%has been assigned.  
%The author list should be included in the draft used for 
%first and second circulation, to allow new members of the collaboration to verify
%that they have been included correctly. Occasionally a misspelled
%name is corrected, or associated institutions become full members.
%Therefore an updated author list will be sent to you after the final
%EB review of the paper.  In case line numbering doesn't work well
%after including the authorlist, try moving the \verb!\bigskip! after
%the last author to a separate line.

\end{document}

%% file: preamble.tex
\usepackage[top=1in, bottom=1.25in, left=1in, right=1in]{geometry}

\usepackage{microtype}
\usepackage{lineno}  % for line numbering during review
\usepackage{xspace} % To avoid problems with missing or double spaces after
\usepackage{caption} %these three command get the figure and table captions automatically small

\usepackage{graphicx}  % to include figures (can also use other packages)
\usepackage{color}
\usepackage{colortbl}
\graphicspath{{./figs/}} % Make Latex search fig subdir for figures
\usepackage{amsmath} % Adds a large collection of math symbols
\usepackage{amssymb}
\usepackage{amsfonts}
\usepackage{upgreek} % Adds in support for greek letters in roman typeset

\newcommand*\patchAmsMathEnvironmentForLineno[1]{%
\expandafter\let\csname old#1\expandafter\endcsname\csname #1\endcsname
\expandafter\let\csname oldend#1\expandafter\endcsname\csname
end#1\endcsname
 \renewenvironment{#1}%
   {\linenomath\csname old#1\endcsname}%
   {\csname oldend#1\endcsname\endlinenomath}%
}
\newcommand*\patchBothAmsMathEnvironmentsForLineno[1]{%
  \patchAmsMathEnvironmentForLineno{#1}%
  \patchAmsMathEnvironmentForLineno{#1*}%
}
\AtBeginDocument{%
\patchBothAmsMathEnvironmentsForLineno{equation}%
\patchBothAmsMathEnvironmentsForLineno{align}%
\patchBothAmsMathEnvironmentsForLineno{flalign}%
\patchBothAmsMathEnvironmentsForLineno{alignat}%
\patchBothAmsMathEnvironmentsForLineno{gather}%
\patchBothAmsMathEnvironmentsForLineno{multline}%
\patchBothAmsMathEnvironmentsForLineno{eqnarray}%
}

\usepackage[pdftex,
            pdfauthor={\paperauthors},
            pdftitle={\paperasciititle},
            pdfkeywords={\paperkeywords}]{hyperref}
\usepackage{hyperxmp}
\hypersetup{
    pdfcopyright={Copyright (C) \papercopyright},
    pdflicenseurl={\paperlicenceurl}
}
\usepackage[colorinlistoftodos,textsize=scriptsize]{todonotes}

\usepackage[bottom,flushmargin,hang,multiple]{footmisc}

\usepackage[all]{hypcap} % Internal hyperlinks to floats.

\input{lhcb-symbols-def} % Add in the predefined LHCb symbols

\hypersetup{
  colorlinks   = true, %Colours links instead of ugly boxes
  urlcolor     = blue, %Colour for external hyperlinks
  linkcolor    = blue, %Colour of internal links
  citecolor    = red   %Colour of citations
}

\ifEnableSectionTOCLinks
    \usepackage[explicit]{titlesec} % to change headings
    
    \let\oldcontentsline\contentsline
    \renewcommand

    \titleformat{\section}{\normalfont\Large\bf}{\hyperlink{tocsection.\thesection}{{\thesection} \parbox[t]{\dimexpr\textwidth-1pc}{#1}}}{1pc}{}

    \titleformat{\subsection}{\normalfont\bf}{\hyperlink{tocsubsection.\thesubsection}{{\thesubsection} \parbox[t]{\dimexpr\textwidth-1pc}{#1}}}{1pc}{}

    \titleformat{name=\section,numberless}[display]{}{}{0pt}{\normalfont\Huge\bfseries #1}
\fi

\usepackage{cite} % Allows for ranges in citations
\usepackage{mciteplus}

%% file: lhcb-symbols-def.tex
\usepackage{xspace} 
\usepackage{upgreek}

\def\lhcb   {\mbox{LHCb}\xspace}

\def\MagUp {\mbox{\em Mag\kern -0.05em Up}\xspace}

\ifthenelse{\boolean{uprightparticles}}%
{

 \def\Ppi         {\ensuremath{\uppi}\xspace}

 \def\PDelta      {\ensuremath{\Delta}\xspace}                 
 \def\PXi         {\ensuremath{\Xi}\xspace}                 
 \def\PLambda     {\ensuremath{\Lambda}\xspace}                 
 \def\PSigma      {\ensuremath{\Sigma}\xspace}                 
 \def\POmega      {\ensuremath{\Omega}\xspace}                 
 \def\PUpsilon    {\ensuremath{\Upsilon}\xspace}
 \let\oldPi\Pi
 \def\PPi         {\ensuremath{\oldPi}\xspace}

 \def\PB      {\ensuremath{\mathrm{B}}\xspace}                 
 \def\PD      {\ensuremath{\mathrm{D}}\xspace}                 

 \def\PK      {\ensuremath{\mathrm{K}}\xspace}                 

 \def\Pc      {\ensuremath{\mathrm{c}}\xspace}

 \def\Pp      {\ensuremath{\mathrm{p}}\xspace}                 

 \def\Ps      {\ensuremath{\mathrm{s}}\xspace}

 \def\thebaroffset{0.0em}
}
{

 \def\Ppi         {\ensuremath{\pi}\xspace}

 \mathchardef\PDelta="7101
 \mathchardef\PXi="7104
 \mathchardef\PLambda="7103
 \mathchardef\PSigma="7106
 \mathchardef\POmega="710A
 \mathchardef\PUpsilon="7107
 \mathchardef\PPi="7105
 \def\PB      {\ensuremath{B}\xspace}                 
 \def\PD      {\ensuremath{D}\xspace}                 

 \def\PK      {\ensuremath{K}\xspace}                 

 \def\Pc      {\ensuremath{c}\xspace}

 \def\Pp      {\ensuremath{p}\xspace}                 

 \def\Ps      {\ensuremath{s}\xspace}

 \def\thebaroffset{0.18em}
}
\newcommand{\offsetoverline}[2][\thebaroffset]{\kern #1\overline{\kern -#1 #2}}%

\makeatletter
\ifcase \@ptsize \relax% 10pt
  \newcommand{\miniscule}{\@setfontsize\miniscule{4}{5}}% \tiny: 5/6
\or% 11pt
  \newcommand{\miniscule}{\@setfontsize\miniscule{5}{6}}% \tiny: 6/7
\or% 12pt
  \newcommand{\miniscule}{\@setfontsize\miniscule{5}{6}}% \tiny: 6/7
\fi
\makeatother

\DeclareRobustCommand{\optbar}[1]{\shortstack{{\miniscule (\rule[.5ex]{1.25em}{.18mm})}
  \\ [-.7ex] $#1$}}

\def\squark    {{\ensuremath{\Ps}}\xspace}

\def\cquark    {{\ensuremath{\Pc}}\xspace}

\def\pion   {{\ensuremath{\Ppi}}\xspace}

\def\pip    {{\ensuremath{\pion^+}}\xspace}

\def\kaon    {{\ensuremath{\PK}}\xspace}
\def\KorKbar {\kern \thebaroffset\optbar{\kern -\thebaroffset \PK}{}\xspace}

\def\Km      {{\ensuremath{\kaon^-}}\xspace}

\def\D       {{\ensuremath{\PD}}\xspace}

\def\DorDbar {\kern \thebaroffset\optbar{\kern -\thebaroffset \PD}\xspace}

\def\Dp      {{\ensuremath{\D^+}}\xspace}
\def\Dm      {{\ensuremath{\D^-}}\xspace}

\def\DpDm    {\ensuremath{\Dp {\kern -0.16em \Dm}}\xspace}

\def\B       {{\ensuremath{\PB}}\xspace}

\def\BorBbar {\kern \thebaroffset\optbar{\kern -\thebaroffset \PB}\xspace}

\def\Bd      {{\ensuremath{\B^0}}\xspace}

\def\BdorBdbar {\kern \thebaroffset\optbar{\kern -\thebaroffset \Bd}\xspace}

\def\Bs      {{\ensuremath{\B^0_\squark}}\xspace}

\def\BsorBsbar {\kern \thebaroffset\optbar{\kern -\thebaroffset \Bs}\xspace}

\def\Y#1S{\ensuremath{\PUpsilon{(#1S)}}\xspace}

\def\proton      {{\ensuremath{\Pp}}\xspace}

\def\Lz          {{\ensuremath{\PLambda}}\xspace}
\def\Lbar        {{\ensuremath{\offsetoverline{\PLambda}}}\xspace}
\def\LorLbar     {\kern \thebaroffset\optbar{\kern -\thebaroffset \PLambda}\xspace}

\def\Lc          {{\ensuremath{\Lz^+_\cquark}}\xspace}
\def\Lcbar       {{\ensuremath{\Lbar{}^-_\cquark}}\xspace}

\newcommand{\decay}[2]{\ensuremath{\mathinner{#1\!\to #2}}\xspace}

\def\to                 {\ensuremath{\rightarrow}\xspace}

\def\AT#1     {\ensuremath{A_{\mathrm{T}}^{#1}}\xspace}           % 2

\def\C#1      {\ensuremath{\mathcal{C}_{#1}}\xspace}                       % 9
\def\Cp#1     {\ensuremath{\mathcal{C}_{#1}^{'}}\xspace}                    % 7
\def\Ceff#1   {\ensuremath{\mathcal{C}_{#1}^{\mathrm{(eff)}}}\xspace}        % 9  
\def\Cpeff#1  {\ensuremath{\mathcal{C}_{#1}^{'\mathrm{(eff)}}}\xspace}       % 7
\def\Ope#1    {\ensuremath{\mathcal{O}_{#1}}\xspace}                       % 2
\def\Opep#1   {\ensuremath{\mathcal{O}_{#1}^{'}}\xspace}                    % 7

\newcommand{\aunit}[1]{\ensuremath{\text{\,#1}}}       
\newcommand{\tev}{\aunit{Te\kern -0.1em V}\xspace}
\newcommand{\gev}{\aunit{Ge\kern -0.1em V}\xspace}
\newcommand{\mev}{\aunit{Me\kern -0.1em V}\xspace}
\newcommand{\kev}{\aunit{ke\kern -0.1em V}\xspace}
\newcommand{\ev}{\aunit{e\kern -0.1em V}\xspace}
 
\newcommand{\mevc}{\ensuremath{\aunit{Me\kern -0.1em V\!/}c}\xspace}
\newcommand{\gevc}{\ensuremath{\aunit{Ge\kern -0.1em V\!/}c}\xspace}
\newcommand{\mevcc}{\ensuremath{\aunit{Me\kern -0.1em V\!/}c^2}\xspace}
\newcommand{\gevcc}{\ensuremath{\aunit{Ge\kern -0.1em V\!/}c^2}\xspace}
\def\mm   {\aunit{mm}\xspace}

\def\nb {\aunit{nb}\xspace}
\def\invnb {\ensuremath{\nb^{-1}}\xspace}

\newcommand{\stat}{\aunit{(stat)}\xspace}
\newcommand{\syst}{\aunit{(syst)}\xspace}

\def\gsim{{~\raise.15em\hbox{$>$}\kern-.85em
          \lower.35em\hbox{$\sim$}~}\xspace}
\def\lsim{{~\raise.15em\hbox{$<$}\kern-.85em
          \lower.35em\hbox{$\sim$}~}\xspace}

\def\sqsnn {\ensuremath{\protect\sqrt{s_{\scriptscriptstyle\text{NN}}}}\xspace}
\def\pt         {\ensuremath{p_{\mathrm{T}}}\xspace}

\def\rad{\aunit{rad}\xspace}

\def\evtgen     {\mbox{\textsc{EvtGen}}\xspace}

\def\geant      {\mbox{\textsc{Geant4}}\xspace}

\def\photos     {\mbox{\textsc{Photos}}\xspace}

\def\pythia     {\mbox{\textsc{Pythia}}\xspace}

\def\hepml      {\mbox{\texttt{hep\_ml}}\xspace}

\def\tell1  {TELL1\xspace}
\def\ukl1   {UKL1\xspace}

\newcommand{\lhcborcid}[1]{\href{https://orcid.org/#1}{\hspace*{0.1em}\raisebox{-0.45ex}{\includegraphics[width=1em]{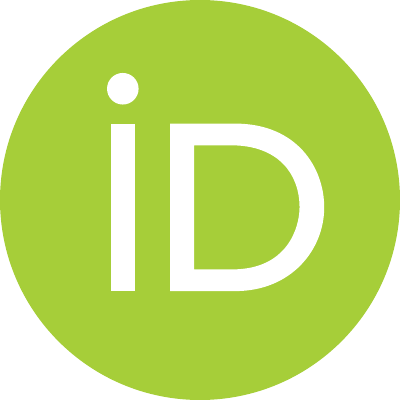}}}}

%% file: thesis-symbols.tex
\newcommand{\Lcpkpi}{\mbox{\decay{\Lc}{p\Km\pip}}\xspace}

\newcommand{\mqpk}{\ensuremath{m^2_{p\Km}}\xspace}
\newcommand{\mqkpi}{\ensuremath{m^2_{\Km\pip}}\xspace}

\newcommand{\xf}{\ensuremath{x_{\rm F}}\xspace}

%% file: title-LHCb-PAPER.tex
% ===============================================================================
% Purpose: LHCb-PAPER journal paper title page template
% Author: 
% Created on: 2010-09-25
% ===============================================================================

%%%%%%%%%%%%%%%%%%%%%%%%%
%%%%%  TITLE PAGE  %%%%%%
%%%%%%%%%%%%%%%%%%%%%%%%%
\begin{titlepage}
\pagenumbering{roman}

% Header ---------------------------------------------------
\vspace*{-1.5cm}
\centerline{\large EUROPEAN ORGANIZATION FOR NUCLEAR RESEARCH (CERN)}
\vspace*{1.5cm}
\noindent
\begin{tabular*}{\linewidth}{lc@{\extracolsep{\fill}}r@{\extracolsep{0pt}}}
\ifthenelse{\boolean{pdflatex}}% Logo format choice
{\vspace*{-1.5cm}\mbox{\!\!\!\includegraphics[width=.14\textwidth]{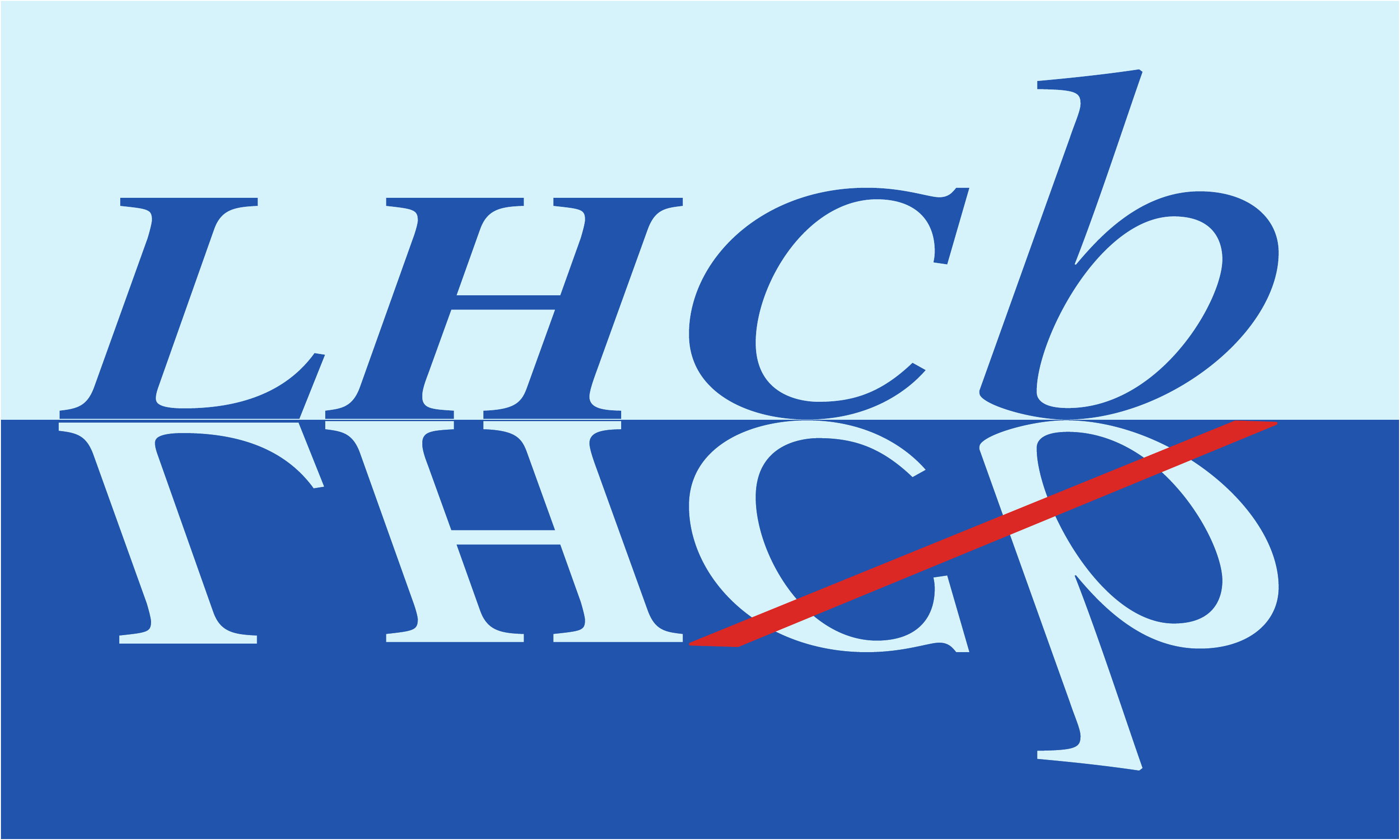}} & &}%
{\vspace*{-1.2cm}\mbox{\!\!\!\includegraphics[width=.12\textwidth]{figs/lhcb-logo.eps}} & &}%
\\
 & & CERN-EP-2026-016 \\  % ID 
 & & LHCb-PAPER-2025-060 \\  % ID 
 & & 28 August 2026 \\ % Date - Can also hardwire e.g.: 23 March 2010
 & & \\
% not in paper \hline
\end{tabular*}

\vspace*{4.0cm}

% Title --------------------------------------------------
{\normalfont\bfseries\boldmath\huge
\begin{center}
% DO NOT EDIT HERE. Instead edit macro in main.tex to keep metadata correct
  \papertitle 
\end{center}
}

\vspace*{2.0cm}

% Authors -------------------------------------------------
\begin{center}
%In the footnote, replace 'paper' by 'Letter' in case of submission to PRL or PLB 
% Edit macro in main.tex to keep metadata correct
\paperauthors\footnote{Authors are listed at the end of this paper.}
\end{center}

\vspace{\fill}

% Abstract -----------------------------------------------
\begin{abstract}
  \noindent
The first measurement of the polarization of charm baryons by the \lhcb experiment recorded in fixed-target mode is presented.
The polarization of $\PLambda_c$ baryons is studied in collisions of protons, at an energy of 2.51\tev, incident on a gaseous target of neon, at a nucleon-nucleon center-of-mass energy of $68.6 \gev$. The world’s first measurement of separate-charge polarizations for $\PLambda^+_c$ and $\offsetoverline{\PLambda}{}^-_c$ baryons is performed, determining
$$ P_{\ensuremath{\PLambda^+_c}} = ( 24 \pm 9 \pm 2 \, )\% , $$ $$ P_{\ensuremath{\offsetoverline{\PLambda}{}^-_c}} = (-8 \pm 12  \pm 3 \, ) \%  , $$
where the first uncertainty is statistical and the second systematic. The polarization is also measured in intervals of baryon transverse momentum and the Feynman-$x$ variable.

\end{abstract}

\vspace*{2.0cm}

\begin{center}
%  Submitted to
Published in Phys.~Rev.~Lett. 137 (2026) 091901
\end{center}

\vspace{\fill}

{\footnotesize 
% Edit macro in main.tex to keep metadata correct
\centerline{\copyright~\papercopyright. \href{\paperlicenceurl}{\paperlicence}.}}
\vspace*{2mm}

\end{titlepage}

%%%%%%%%%%%%%%%%%%%%%%%%%%%%%%%%
%%%%%  EOD OF TITLE PAGE  %%%%%%
%%%%%%%%%%%%%%%%%%%%%%%%%%%%%%%%

%  empty page follows the title page ----
\newpage
\setcounter{page}{2}
\mbox{~}

%% file: body.tex
Charm baryon polarization is a unique probe in high-energy physics, giving preferential access to heavy-quark spin-dependent physics in its production and subsequent hadronization.
The heavy quark drives baryon properties, including polarization, since its mass is much larger than the typical energy scale of strong nuclear interactions~\cite{Mannel:1991bs,Falk:1993rf,Galanti:2015pqa}.
This is in contrast with lighter baryons, whose polarization is not directly related to the spins of their constituents~\cite{Ashman:1987hv,deFlorian:2008mr,Adolph:2015saz}. Strong interaction polarization arises mainly from spin-orbit correlations and final-state interactions of quarks, leading to different baryon and antibaryon polarization mechanisms due to the distinct role of valence and sea quarks. Polarization measurements provide tests for QCD dynamics in the nonperturbative regime, helping to distinguish among various models~\cite{Andersson:1979wj,Szwed:1981rr,DeGrand:1981pe,Dharmaratna:1989jr,Barni:1991qn,Soffer:1991am,Hama:1993aq,Dharmaratna:1996xd,Troshin:1996hd,Liang:1997rt,Felix:1999tf,Anselmino:2000vs,Kanazawa:2000cx,Anselmino:2001js,Boer:2007nh,Zhou:2008fb,Kanazawa:2015jxa,Koike:2015zya,Koike:2017fxr,Gamberg:2018fwy,DAlesio:2020wjq}.

Experimentally, charm baryon polarization arising from strong interactions remains a poorly explored topic.
The averaged \Lc and \Lcbar polarization has only been measured in $\pi$-nucleus fixed-target collisions by the NA32~\cite{Jezabek:1992ke} and E791~\cite{Aitala:1999uq} experiments, for beam energies of 230 and 500\gev, respectively.
Both indicated a trend of increasing negative polarization with larger transverse momentum, \pt, though with large uncertainties. A similar \pt trend was also observed for strange \Lz baryons in fixed-target experiments~\cite{Bunce:1976yb,Heller:1978ty,Lundberg:1989hw,Yuldashev:1990az, RAMBERG1994403,na48,HADES:2014ttv}. In addition, a trend of increasing polarization with $|\xf|$ has been observed for \Lz baryons~\cite{Lundberg:1989hw, RAMBERG1994403,na48, Abt_2006, LHCb-PAPER-2024-009}, where $\xf\equiv\frac{2p_{\rm L}}{\sqsnn}$ is the Feynman-$x$ variable, $p_{\rm L}$ is the baryon momentum component parallel to the beam axis in the center-of-mass frame, and \sqsnn the nucleon-nucleon center-of-mass energy.
However, no charm baryon polarization measurement as a function of \xf has been performed to date. 
The E791 results have been used to develop a QCD-based hybrid model of inclusively produced charm baryons reproducing the \pt dependence~\cite{Goldstein:1999jr}. No further theoretical studies of charm baryon polarization have been done since then.
 
In this Letter, the first measurement of the polarization of charm baryons by the \lhcb experiment in its fixed-target mode is presented. The polarization of $\PLambda_c$ baryons is studied in collisions of protons, with an energy of 2.51\tev, incident on a gaseous target of neon, taken in 2017. This corresponds to an energy $\sqsnn = 68.6 \gev$, an energy regime not previously tested for polarization studies. The dataset corresponds to an integrated luminosity of $21.7\invnb$. This study contributes to the largely unexplored field of charm baryon polarization measurements, providing new inputs to test low-energy QCD models for heavy-baryon spin physics and the kinematic dependencies observed in hyperon polarization.
In addition, it improves different aspects of the previous polarization measurements, as described in the following.

The polarization is measured for the first time separately for the \Lc and \Lcbar hadrons, to probe their different production mechanisms. Their decays are reconstructed in the \Lcpkpi final state, with charge-conjugate states implied unless specified. The polarization is extracted using a detailed amplitude model~\cite{Marangotto:2020ead}. The model is based on the helicity formalism, and, in contrast to what has been done previously, it ensures that the spin states of the final particles have the same exact definition for all the intermediate resonant contributions~\cite{Marangotto:2019ucc}, permitting a correct description of interference effects. The amplitude model permits to combine the different sensitivities given by the many resonant contributions in the whole \Lcpkpi phase space to achieve the best precision on the polarization measurement. Indeed, the \Lcpkpi decay is the main \Lc decay channel and it has a large sensitivity to polarization, $\sqrt{3}S \approx 0.662$~\cite{LHCb-PAPER-2022-002}, where $\sqrt{3}S$ is a quantity equivalent to two-body decay asymmetry parameters, linked to the amount of parity-violation in the decay.
The present study is based on a novel technique for a polarization measurement, where the amplitude model is determined from a different signal sample with larger statistics. The model description and parameters have been previously determined using a sample of $4\times 10^5$ \Lcpkpi candidates recorded in semileptonic $b$-hadron decays from \proton\proton collisions by the LHCb experiment~\cite{LHCb-PAPER-2022-002}. 
The E791 analysis~\cite{Aitala:1999uq} has previously explored a helicity-formalism-based technique, but the model did not consider the issue of the matching of final particle spin states, and was developed on a far smaller dataset.
In addition, this Letter measures the polarization as a function of \pt and \xf in order to probe kinematic polarization dependencies, testing the trends suggested by previous charm and \Lz measurements~\cite{Bunce:1976yb,Heller:1978ty,Lundberg:1989hw,Yuldashev:1990az, RAMBERG1994403,na48, HADES:2014ttv, Abt_2006, LHCb-PAPER-2024-009}.
A significant \Lc polarization in fixed-target collisions at LHC beam energies would enable the measurement of charm-baryon electric- and magnetic-dipole moments via spin precession~\cite{Botella:2016ksl,Bagli:2017foe,Fomin:2017ltw,Marangotto:2713231,Fomin:2019wuw,Mirarchi:2019vqi,Aiola:2020yam,Neri:2024rjv,Akiba:2905467}.

% Detector & Simulation

The LHCb detector is a single-arm forward spectrometer covering the pseudorapidity range $2 < \eta < 5$, described in detail in Refs.~\cite{LHCb-DP-2008-001,LHCb-DP-2014-002}.
The detector used to collect the data analyzed in this paper includes: a silicon-strip vertex locator (VELO), three tracking stations of silicon-strip detectors and straw drift tubes,
two ring-imaging Cherenkov detectors (RICH) that are able to discriminate between different species of charged hadrons, a calorimeter system consisting of scintillating-pad and preshower detectors, electromagnetic and hadronic calorimeters, and a muon detector composed of alternating layers of iron and multiwire proportional chambers.
The System for Measuring the Overlap with Gas (SMOG)~\cite{FerroLuzzi:2005em} enables the injection of gases with a pressure of $\mathcal{O}(10^{-7})$ mbar in the beam pipe section inside the VELO, allowing LHCb to effectively operate as a fixed-target experiment. The SMOG system thus provides a unique opportunity to study proton-nucleus and nucleus-nucleus collisions with various gaseous targets using the LHC beams. In this configuration, the LHCb acceptance extends to the negative center-of-mass rapidity hemisphere $-2.5 < y^* < 0$, due to the boost induced by the high-energy proton beam.

Fixed-target $p$Ne collision events were only recorded when a bunch in the beam pointing towards the LHCb detector crossed the interaction region without a corresponding bunch in the beam pointing in the opposite direction. In order to suppress background $pp$ collision events, the $z$ coordinate of the $p$Ne primary vertex (PV) is required to lie outside the region $z_{\rm{PV}} \in [-100,100]\mm$, where $z_{\rm{PV}} = 0\mm$ is at the center of the $pp$ interaction region, and the standard deviation of the distribution is about $50\mm$. Events lying inside the region are retained only if no hits are detected in the VELO stations upstream of the interaction region.

Efficiencies are estimated using samples of simulated \Lcpkpi decays from $p$Ne collisions.
Unpolarized \Lcpkpi decays are uniformly generated over the phase space using \pythia~\cite{Sjostrand:2007gs} with a specific LHCb configuration~\cite{LHCb-PROC-2010-056} and with colliding-proton beam momentum equal to the momentum per nucleon of the beam and target in the center-of-mass frame. The decays of unstable particles are described by \evtgen~\cite{Lange:2001uf}, in which final-state radiation is generated using \photos~\cite{davidson2015photos}.
The four-momenta of signal decay particles are extracted and embedded into $p$Ne events generated with the \textsc{Epos} event generator~\cite{Pierog:2013ria}, which models the underlying inelastic collision.
The interaction of the generated particles with the detector and its response are implemented using the \geant toolkit~\cite{Agostinelli:2002hh,Allison:2006ve} as described in Ref.~\cite{LHCb-PROC-2011-006}.
Simulated distributions are validated on data, correcting observed differences using the gradient-boosted decision tree weighting technique from the \hepml package~\cite{Rogozhnikov:2016bdp}.

%Selection

The \Lcpkpi decays are reconstructed from combinations of three charged tracks forming a vertex separated from the PV. Only events with a single reconstructed PV are retained for further analysis. Charged tracks are required to be identified as hadrons by loose particle-identification criteria, and must have $\pt>200\mev$, momentum $p>3\gev$ and be significantly displaced from the PV.
Natural units with $\hbar = c = 1$ are used throughout.
All reconstructed tracks and vertices are required to have good quality fits.
Further selection criteria are imposed to reduce the combinatorial background arising from random combinations of tracks. These require each track to be within the \lhcb detector pseudorapidity acceptance, $2<\eta<5$, and to pass particle-identification criteria obtained from neural network classifiers, which combine information from the different subdetectors~\cite{LHCb-DP-2018-001}. Reconstructed \Lc candidates are required to have a decay vertex displaced in the direction of the momentum, a small impact parameter with respect to the PV and a maximum value of the angle between the momentum vector and the flight direction of $0.25 \rad$.
No evidence of misidentified, wrongly reconstructed heavy-hadron decays or candidates originating from beauty-hadron decays is found.

A boosted decision tree (BDT) classifier~\cite{Breiman} is used to further reduce the combinatorial background. The classifier is trained on the \Lcpkpi simulation sample as a proxy for the signal, and on data candidates in invariant-mass sidebands as a proxy for the background, with sidebands defined as $40 < |m(\proton\Km\pip) - m(\Lc)| <100\mev$, and $m(\Lc)$ the known \Lc mass~\cite{PDG2024}. The $k$-folding technique is applied~\cite{kFold}  to limit the possible bias due to the limited signal proxy yield.
The BDT exploits quantities related to the decay topology and particle identification, as well as \pt and momentum of the \Lc candidates. A selection requirement on the BDT classifier output is chosen by optimizing the figure of merit $N^2/(N+B)^{3/2}$, where $N$ and $B$ are the fitted signal and background yields in the signal region, defined as that for which $m(\proton\Km\pip)$ is within $\pm 15\mev$ of the known \Lc mass~\cite{PDG2024}.

\begin{figure}
\centering
\includegraphics[width=0.49\textwidth]{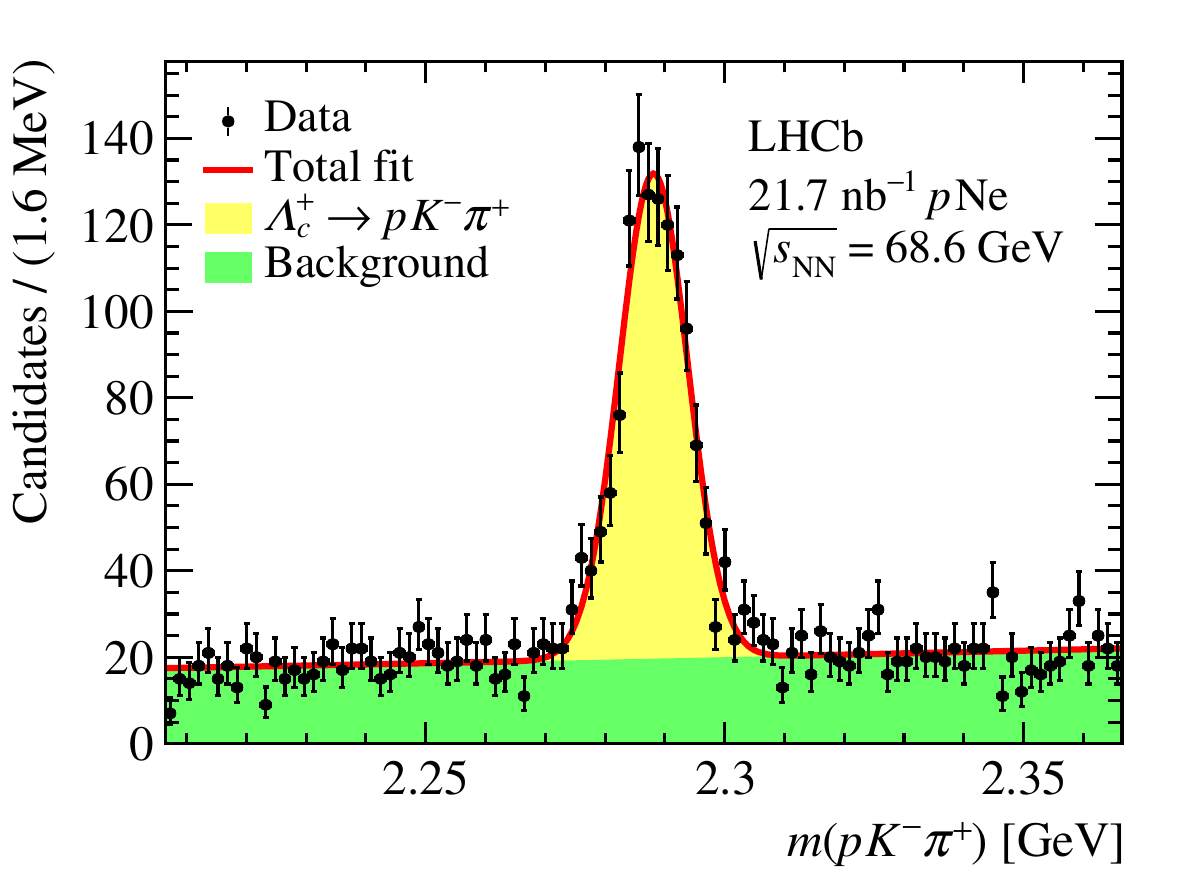}
\includegraphics[width=0.49\textwidth]{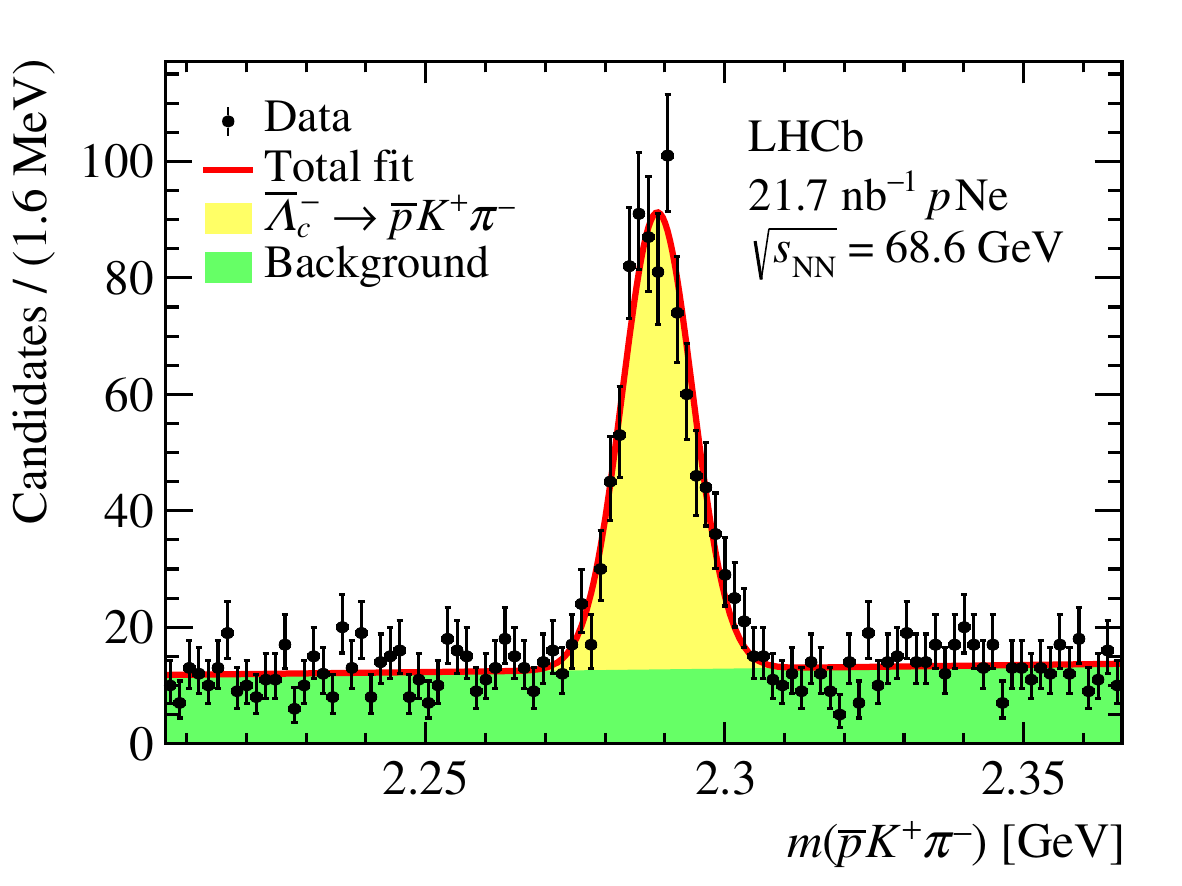}
\caption{Distributions of the $pK^-\pi^+$ invariant mass of selected (left) \Lc and (right) \Lcbar candidates.
The results from the fit described in the text are also shown.
\label{fig:fit}}
\end{figure}

The distributions of the invariant mass, $m(\proton\Km\pip)$, of selected \Lc and \Lcbar candidates are shown in Fig.~\ref{fig:fit}.
Extended unbinned maximum-likelihood fits are performed to determine the signal and background yields after the selection. The signal shape is modeled with a Gaussian probability density function, while an exponential model is used to describe the combinatorial background.
Candidates within the signal region are retained for the polarization measurement. The \Lc and \Lcbar signal yields are respectively determined to be $997 \pm 39$ and $710 \pm 32$, with background fractions of $(27.0 \pm 1.1)\%$ and $(25.1 \pm 1.2)\%$, where the uncertainties are statistical only. The measured signal yields are consistent with a substantial particle-antiparticle production asymmetry, which is expected in proton-induced fixed-target collisions at these center-of-mass energies.

% Polarization strategy

The baryon polarization is measured with respect to the production plane formed by the beam direction, taken as the $z$-axis of the polarization system, and the baryon momentum, whose vector product defines the transverse polarization component, constituting the $y$ axis. This is the only component measured in this study, as the others are forbidden by parity conservation in strong interactions.
The five variables describing the \Lcpkpi decay phase space are chosen to be the squared invariant masses \mqpk and \mqkpi, and quantities that describe the decay orientation with respect to the polarization system $\cos\theta_p,\phi_p$ and $\chi$. The angles $\theta_p$ and $\phi_p$ are the polar and azimuthal angles of the proton momentum, while $\chi$ is the angle between the plane formed by the proton momentum and the beam direction, and the plane formed by the kaon and pion momenta, where the momenta are expressed in the \Lc rest frame. 
The \Lc and \Lcbar transverse polarizations are measured by independent maximum-likelihood fits to phase-space distributions fixing the amplitude model to that previously determined by the LHCb collaboration~\cite{LHCb-PAPER-2022-002}, with the other polarization components fixed to zero.
Nonuniformity of the detector acceptance across the \Lcpkpi phase space is incorporated using simulation following the method outlined in Ref.~\cite{LHCb-PAPER-2022-002}, while the background contribution is accounted for using factorized Legendre-polynomial expansions, derived from the mass sidebands in data.
The fit implementation is based on a version of the \mbox{\textsc{TensorFlowAnalysis}} package~\cite{TFA} adapted to three-body amplitudes in five-dimensional phase-space fits~\cite{Marangotto:2713231}.

% Uncertainties

Statistical uncertainties are determined using a nonparametric bootstrapping technique from the standard deviation of 1000 pseudoexperiments generated from the data fit result.
The systematic uncertainty associated to the \Lcpkpi amplitude model uncertainty is estimated using the alternative models described in Ref.~\cite{LHCb-PAPER-2022-002}, representing a variety of descriptions for the resonant and spin structure of the decay.
The systematic uncertainty associated to the background description includes the contribution arising from the background fraction, varied according to its uncertainty, and from its parametrization, which is estimated by considering alternatives derived from bootstrapped data samples extracted from the mass sidebands.
The uncertainty associated to the description of detector efficiency effects is estimated by varying the calibration samples via bootstrapping and changing the parameters of the decision tree employed. A negligible fit bias is observed.
Systematic uncertainties for each contribution are reported in Table~\ref{tab:syst_pol}.
\begin{table}
\centering
\caption{Systematic uncertainty contributions to the \Lc and \Lcbar polarization measurements.\label{tab:syst_pol}}
\begin{tabular}{lcccc}
\toprule
& Total & Model & Background & Efficiency\\
\midrule
$P_\Lc$ & 0.022 & 0.016 & 0.009 & 0.014 \\
$P_\Lcbar$ & 0.031 & 0.016 & 0.018 & 0.020 \\
\bottomrule
\end{tabular}
\end{table}
The stability of the measurement is tested by performing a fit in which all components of the polarization vector are allowed to vary independently for both the \Lc and \Lcbar datasets. The polarization components within the production plane are found to be compatible with zero as required by parity symmetry, while no significant shift is seen for the transverse polarization component.

\begin{figure}[t]
\centering
\includegraphics[width=0.95\textwidth]{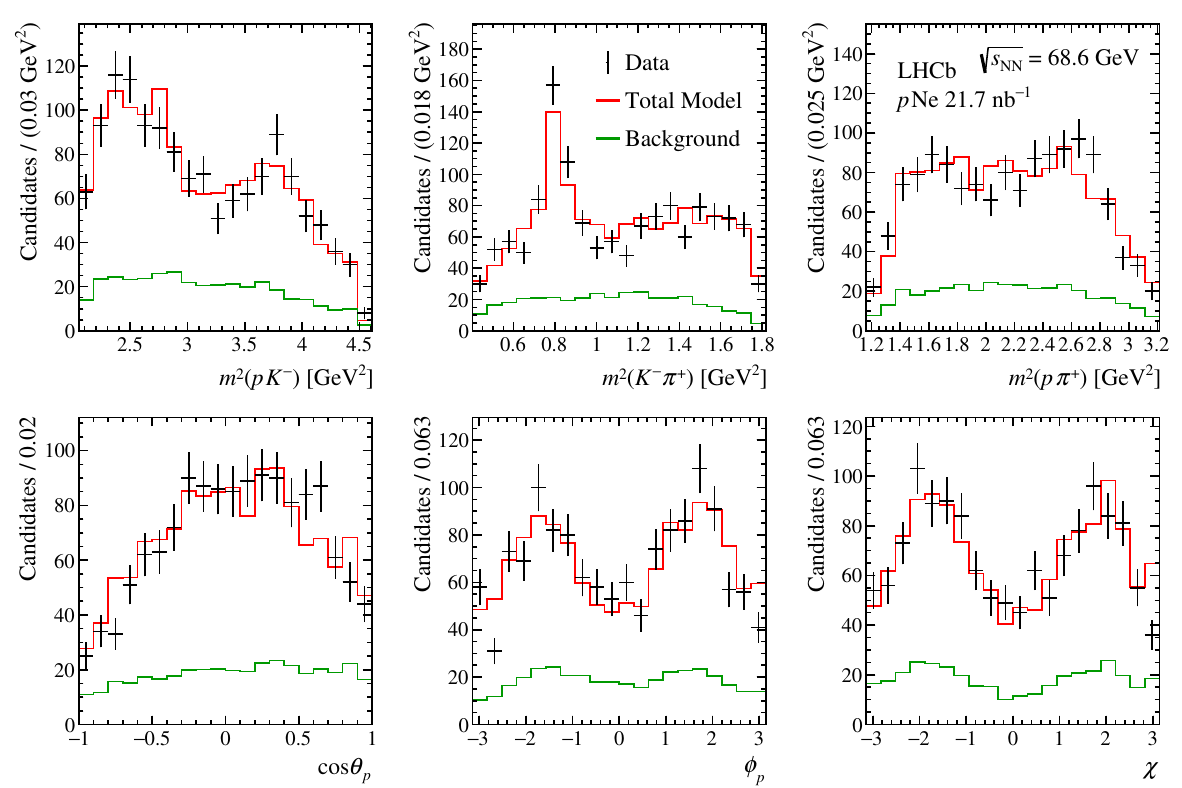}
\caption{Comparisons of the phase-space projections between the data and the fitted model, for the \Lc dataset. \label{fig:pol_fit_Lc}}
\end{figure}

\begin{figure}[tb]
\centering
\includegraphics[width=0.95\textwidth]{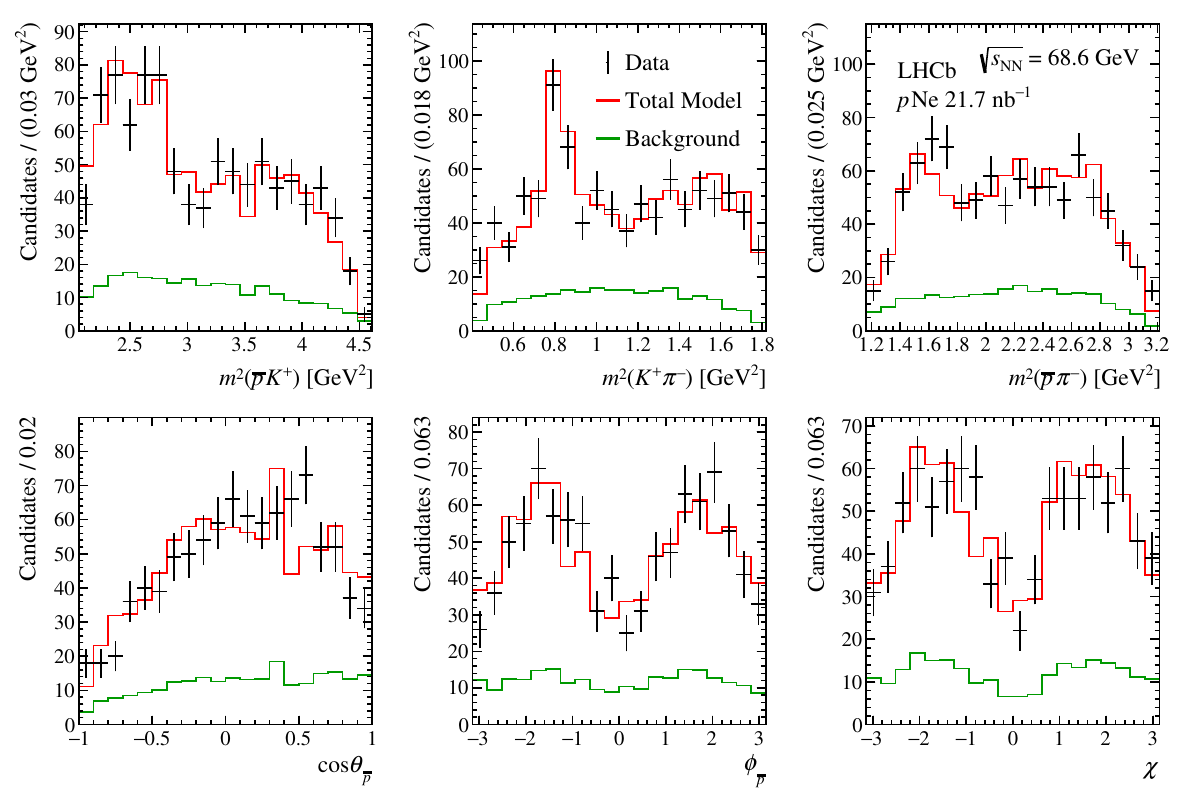}
\caption{Comparisons of the phase-space projections between the data and the fitted model, for the \Lcbar dataset. \label{fig:pol_fit_Lcbar}}
\end{figure}

% Results
Phase-space distribution comparisons between the data and the fitted model are shown in Figs.~\ref{fig:pol_fit_Lc} and~\ref{fig:pol_fit_Lcbar} for the \Lc and \Lcbar datasets, respectively.
The polarizations for the \Lc and \Lcbar datasets are determined to be
\begin{align*}
P_\Lc &= ( 24.4 \pm 9.5 \, (\rm{stat}) \pm 2.2 \, (\rm{syst}) \, )\% ,\\
P_\Lcbar &= (-8 \pm 12 \, (\rm{stat}) \pm 3 \,(\rm{syst}) \, ) \%.
\end{align*}
An indication of a sizable \Lc polarization is present, having a statistical significance of $2.4\,\sigma$, evaluated using pseudoexperiments. Compared with previous charm and \Lz measurements, an opposite polarization is seen due to the negative \xf coverage of the LHCb SMOG data, and given $P(-\xf) = -P(\xf)$.
The \Lcbar polarization is compatible with zero within the uncertainty.

The differential polarization measurement is performed by splitting both datasets into three intervals of \pt and \xf, chosen in order to have a comparable number of candidates in each interval. Numerical polarization results are reported in Table~\ref{tab:pol_results_diff}, and shown in the End Matter, which also provides background-subtracted \pt and \xf distributions.
The larger \Lc polarization value in the highest \pt bin suggests a possible trend of increasing polarization with \pt, as seen in previous charm~\cite{Jezabek:1992ke,Aitala:1999uq} and \Lz measurements~\cite{Bunce:1976yb,Heller:1978ty,Lundberg:1989hw,Yuldashev:1990az, RAMBERG1994403,na48,Abt_2006,HADES:2014ttv,  LHCb-PAPER-2024-009}, although the statistical significance is inconclusive.
No clear kinematic dependencies are visible for \Lcbar polarization.

\begin{table}[b]
\centering
\caption{Results of the differential polarization measurement.\label{tab:pol_results_diff}}
\begin{tabular}{lr@{\:$\pm$\:}c@{\:$\pm$\:}lr@{\:$\pm$\:}c@{\:$\pm$\:}l}
\toprule
& $P_\Lc$ & \stat & \syst [\%] & $P_\Lcbar$  & \stat & \syst [\%]\\
\midrule
$\pt<2.8 \gev$ & 17  & 17  & 4 & $-2$  & 20  & 4 \\
$2.8\leq \pt <4 \gev$ & 18  & 14  & 4 & $-6$  & 17  & 5\\
$\pt\geq 4 \gev$ & 45  & 17  & 4  & $-27$  & 22  & 14\\
\midrule
$\xf<-0.15$ & 41  & 17  & 5 & $-15$  & 21  & 6 \\
$-0.15\leq \xf<-0.06$ & 9  & 14  & 3 & $-13$  & 18  & 3\\
$\xf\geq -0.06$ & 42  & 18  & 6 & 2  & 20  & 8\\
\bottomrule
\end{tabular}
\end{table}

In summary, the first measurement of the polarization of charm baryons by the \lhcb experiment in its fixed-target mode is reported. It provides the first separate \Lc and \Lcbar polarization measurements, with the first investigation of charm baryon \xf polarization dependence, using a novel helicity-formalism-based amplitude model technique. The \Lc and \Lcbar polarization are \mbox{$P_\Lc = ( 24 \pm 9 \, (\rm{stat}) \pm 2 \, (\rm{syst}) \, )\%$} and \mbox{$P_\Lcbar = (-8 \pm 12 \, (\rm{stat}) \pm 3 \,(\rm{syst}) \, ) \%$}. The differential measurement hints at possible trends of increasing \Lc polarization with \pt as in previous measurements. These results are a first step towards the understanding of charm baryon polarization mechanisms, contributing to the almost empty field of heavy baryon polarization measurements, and paving the way for precise investigations with larger datasets.

%% file: acknowledgements.tex
\section*{Acknowledgements}
%
% These Acknowledgements valid from 12/09/25
%
\noindent We express our gratitude to our colleagues in the CERN
accelerator departments for the excellent performance of the LHC. We
thank the technical and administrative staff at the LHCb
institutes.
We acknowledge support from CERN and from the national agencies:
ARC (Australia);
CAPES, CNPq, FAPERJ and FINEP (Brazil); 
MOST and NSFC (China); 
CNRS/IN2P3 (France); 
BMFTR, DFG and MPG (Germany);
INFN (Italy); 
NWO (Netherlands); 
MNiSW and NCN (Poland); 
MCID/IFA (Romania); 
%MSHE (Russia); 
MICIU and AEI (Spain);
SNSF and SER (Switzerland); 
NASU (Ukraine); 
STFC (United Kingdom); 
DOE NP and NSF (USA).
%%%%%%%%%%%%%%%%%%%%%%%%%%%%%%%%%%%%%%%%%%%%%
We acknowledge the computing resources that are provided by ARDC (Australia), 
CBPF (Brazil),
CERN, 
IHEP and LZU (China),
IN2P3 (France), 
KIT and DESY (Germany), 
INFN (Italy), 
SURF (Netherlands),
Polish WLCG (Poland),
IFIN-HH (Romania), 
%RRCKI and Yandex LLC (Russia), 
PIC (Spain), CSCS (Switzerland), 
and GridPP (United Kingdom).
%%%%%%%%%%%%%%%%%%%%%%%%%%%%%%%%%%%%%%%%%%
We are indebted to the communities behind the multiple open-source
software packages on which we depend.
%%%%%%%%%%%%%%%%%%%%%%%%%%%%%%%%%%%%%%%%%%
Individual groups or members have received support from
%ARC and ARDC (Australia); % moved to national 16/01
Key Research Program of Frontier Sciences of CAS, CAS PIFI, CAS CCEPP, 
%Fundamental Research Funds for the Central Universities,  and Sci.\
%\& Tech.\ Program of Guangzhou (China); Removed 24/11/25
Minciencias (Colombia);
EPLANET, Marie Sk\l{}odowska-Curie Actions, ERC and NextGenerationEU (European Union);
A*MIDEX, ANR, IPhU and Labex P2IO, and R\'{e}gion Auvergne-Rh\^{o}ne-Alpes (France);
%RFBR, RSF and Yandex LLC (Russia);
Alexander-von-Humboldt Foundation (Germany);
ICSC (Italy); 
%GVA, XuntaGal, GENCAT, Inditex, InTalent and Prog.~Atracci\'on Talento, CM (Spain);
Severo Ochoa and Mar\'ia de Maeztu Units of Excellence, GVA, XuntaGal, GENCAT, InTalent-Inditex and Prog.~Atracci\'on Talento CM (Spain);
SRC (Sweden);
the Leverhulme Trust, the Royal Society and UKRI (United Kingdom).

%% file: supplementary-app.tex
\clearpage

\section*{End matter}
\label{sec:Supplementary-App}

\begin{figure}[h]
\centering
\includegraphics[width=0.6\textwidth]{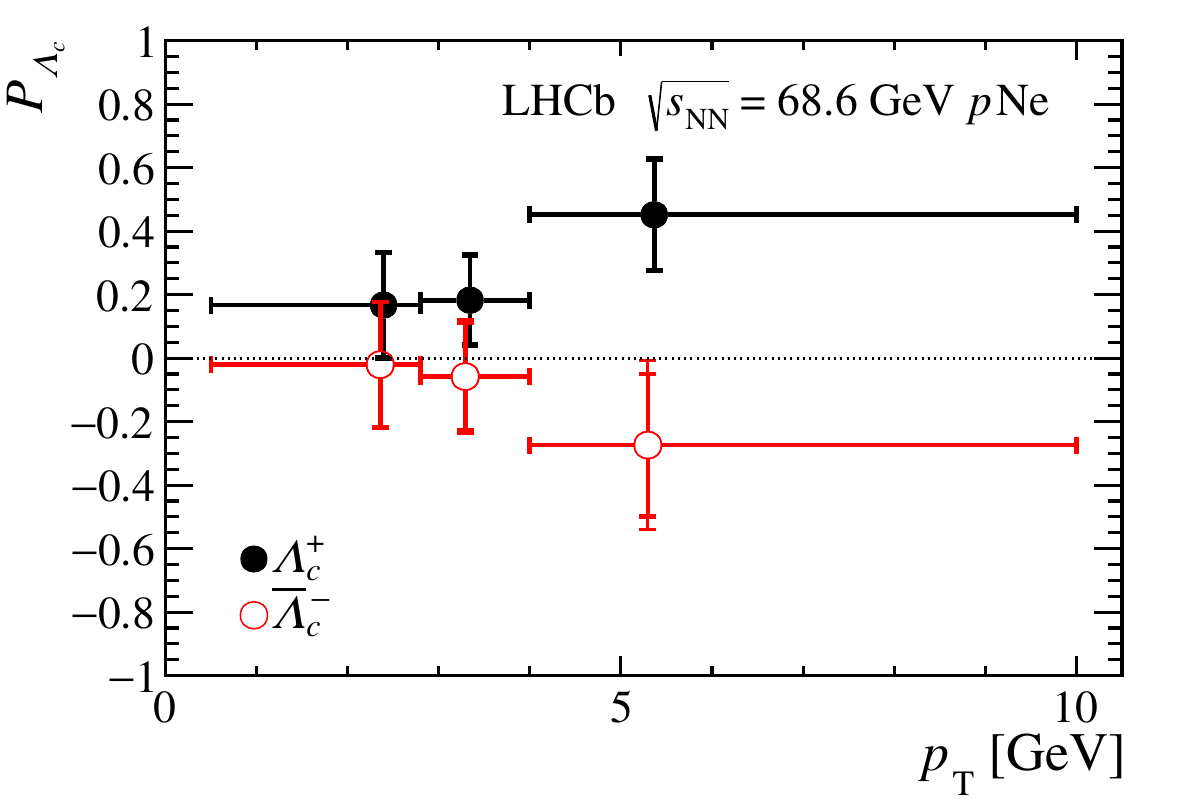}
\caption{Differential polarization measurements in \pt intervals, with (thick line) statistical and (thin line) total uncertainties reported.\label{fig:pol_diff_pt}}
\end{figure}

\begin{figure}[h]
\centering
\includegraphics[width=0.6\textwidth]{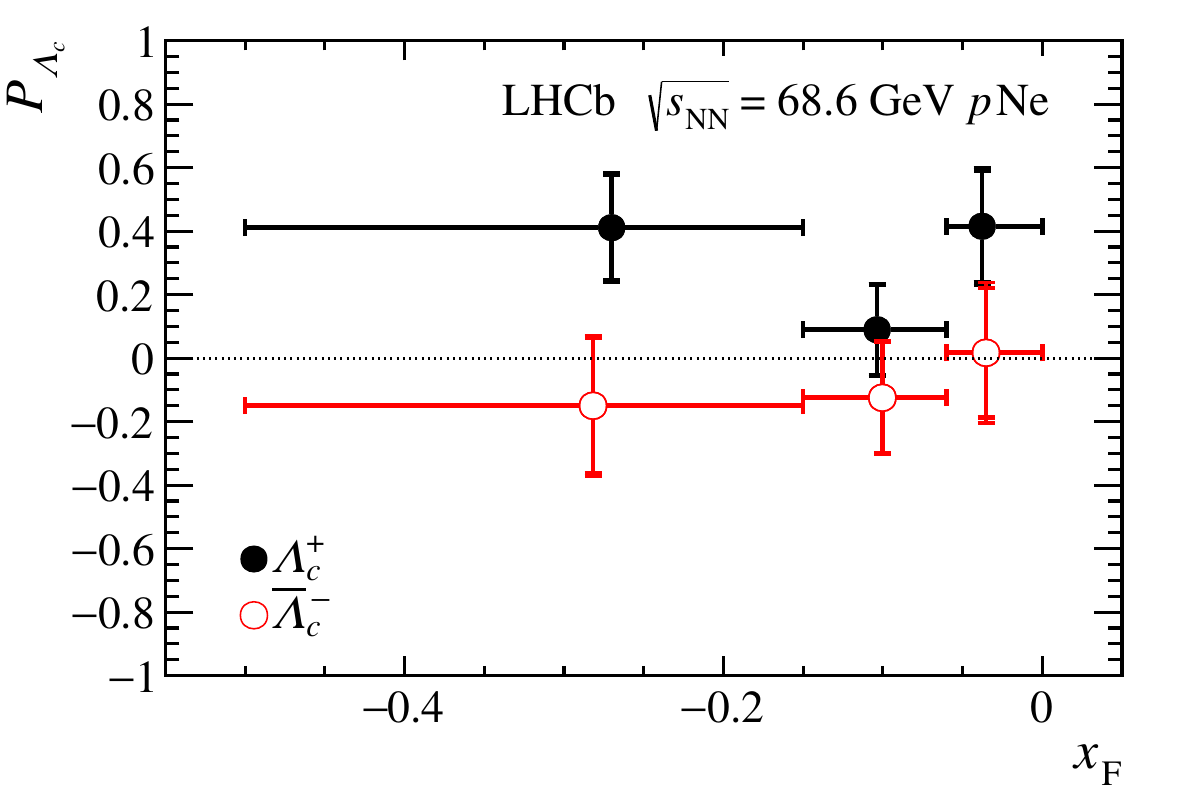}
\caption{Differential polarization measurements in \xf intervals, with (thick line) statistical and (thin line) total uncertainties reported.\label{fig:pol_diff_xf}}
\end{figure}

\begin{figure}[h]
\centering
\includegraphics[width=0.45\textwidth]{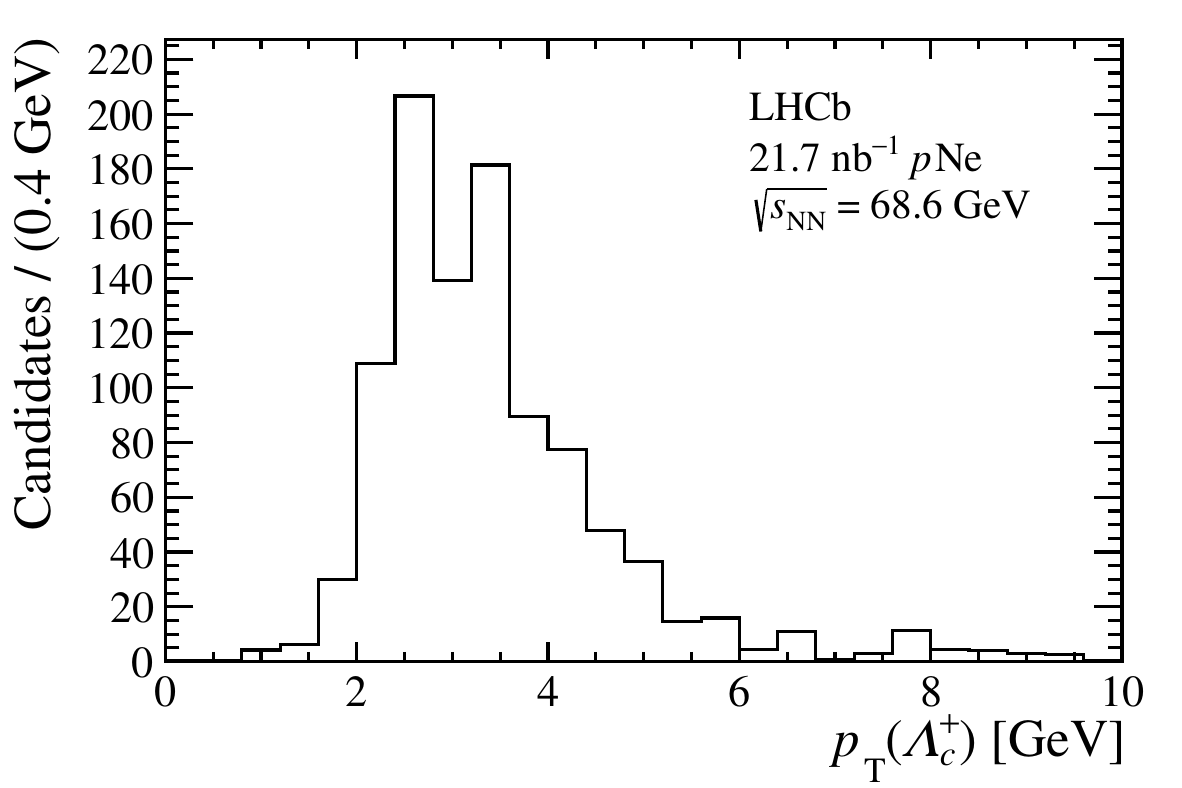}
\includegraphics[width=0.45\textwidth]{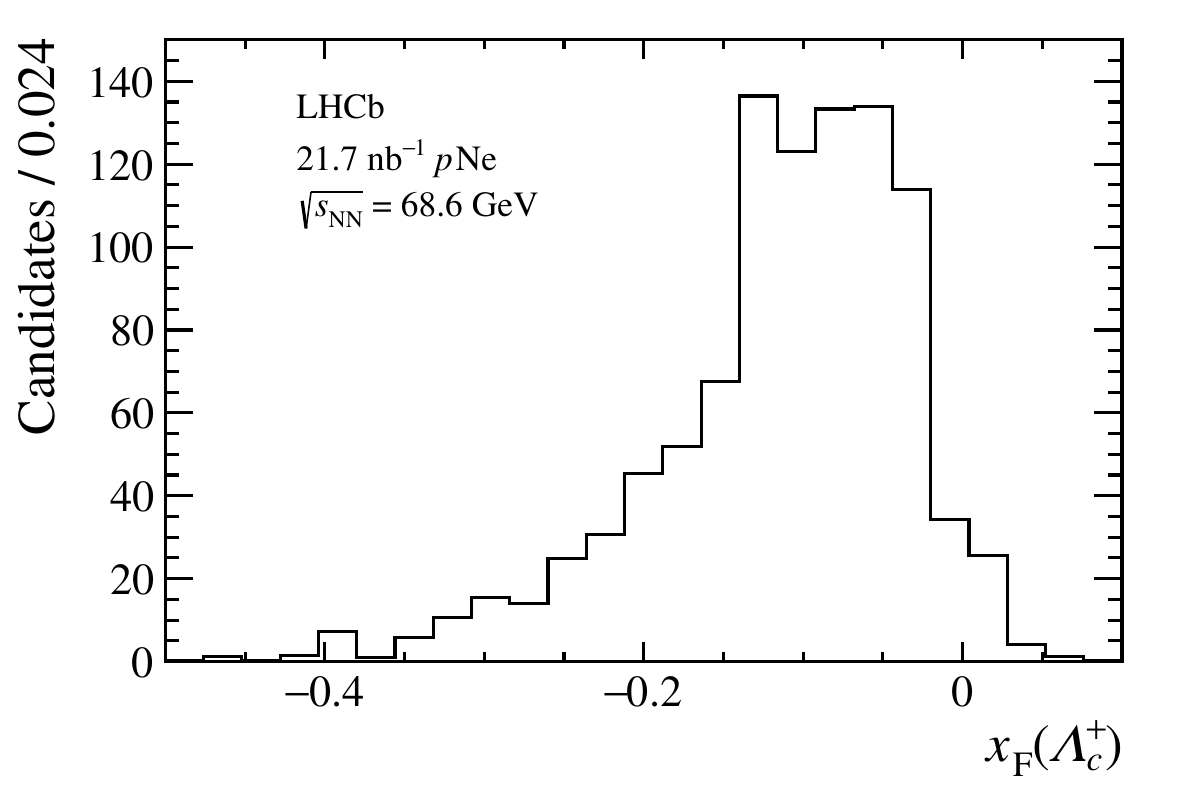}\\
\includegraphics[width=0.45\textwidth]{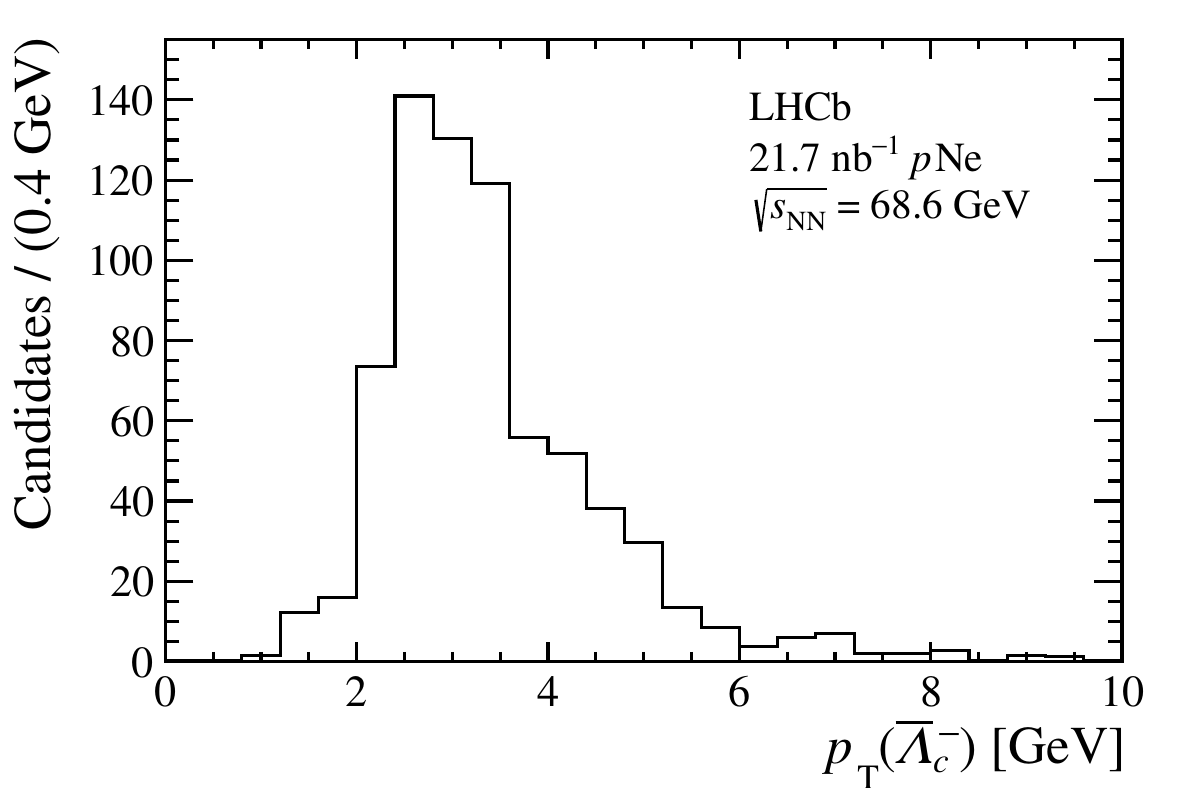}
\includegraphics[width=0.45\textwidth]{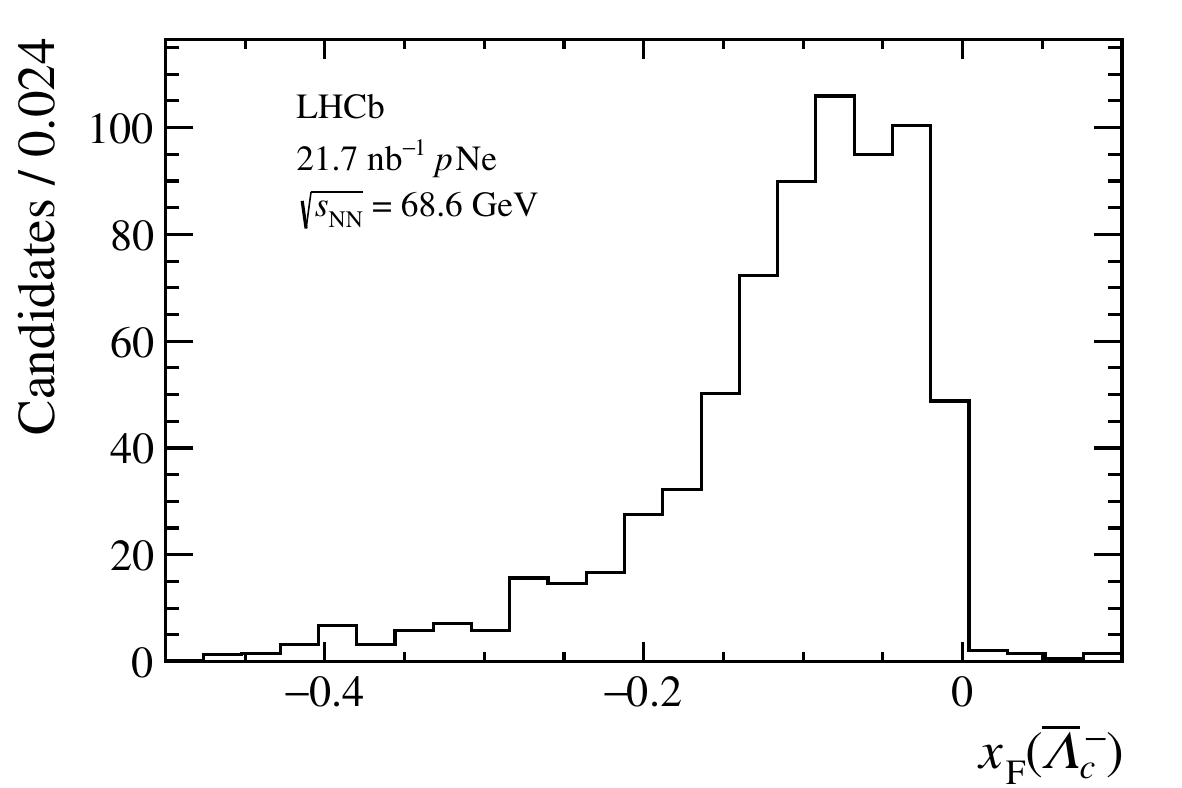}
\caption{Background-subtracted (left) \pt and (right) \xf distributions of \Lc (top) and \Lcbar (bottom) baryon candidates. \label{fig:kin_distr}}
\end{figure}

The differential polarization measurement results are shown in Figs.~\ref{fig:pol_diff_pt} and~\ref{fig:pol_diff_xf}. Systematic uncertainties separated by contribution are reported in Table~\ref{tab:syst_unc_diff}. Background-subtracted \pt and \xf kinematic distributions are reported in Fig.~\ref{fig:kin_distr}, separately for the \Lc and \Lcbar candidates. The average \pt and \xf values, for the full datasets and for each differential polarization fit bin, are reported in Tables~\ref{tab:pt_distr} and~\ref{tab:xf_distr}, respectively.

\begin{table}
\centering
\caption{Systematic uncertainty contributions to the differential polarization measurement.\label{tab:syst_unc_diff}}
\begin{tabular}{llcccc}
\toprule
&  & Total & Model & Background & Efficiency\\
\midrule
$P_\Lc$ & $\pt<2.8 \gev$ & 0.038 & 0.030 & \phantom{$<$}0.001 & 0.024 \\
& $2.8\leq \pt <4 \gev$ & 0.042 & 0.009 & \phantom{$<$}0.002 & 0.041\\
& $\pt\geq 4 \gev$ & 0.041 & 0.030 & \phantom{$<$}0.003 & 0.028\\
& $\xf<-0.15$ & 0.049 & 0.020 & \phantom{$<$}0.002 & 0.045 \\
& $-0.15\leq \xf<-0.06$ & 0.032 & 0.022 & \phantom{$<$}0.002 & 0.024 \\
& $\xf\geq -0.06$ & 0.061 & 0.043 & \phantom{$<$}0.008 & 0.042\\
\midrule
$P_\Lcbar$ & $\pt<2.8 \gev$ & 0.036 & 0.025 & \phantom{$<$}0.003 & 0.025 \\
& $2.8\leq \pt <4 \gev$ & 0.054 & 0.025 & \phantom{$<$}0.002 & 0.048\\
& $\pt\geq 4 \gev$ & 0.14\phantom{0} & 0.03\phantom{0} & $<$0.01\phantom{0} & 0.14\phantom{0}\\
& $\xf<-0.15$ & 0.057 & 0.036 & \phantom{$<$}0.002 & 0.045\\
& $-0.15\leq \xf<-0.06$ & 0.032 & 0.019 & \phantom{$<$}0.003 & 0.025\\
& $\xf\geq -0.06$ & 0.085 & 0.017 & \phantom{$<$}0.001 & 0.083\\
\bottomrule
\end{tabular}
\end{table}

\begin{table}[b]
\centering
\caption{Average \pt values for the full dataset and the differential fit bins. \label{tab:pt_distr}}
\begin{tabular}{lcc}
\toprule
Bin & $\langle \pt \rangle (\Lc) [\gev]$ & $\langle \pt \rangle (\Lcbar) [\gev]$ \\
\midrule
Full range & 3.50 & 3.46 \\
$\pt < 2.8 \gev$ & 2.40 & 2.36 \\
$2.8 \leq \pt < 4 \gev$ & 3.34 & 3.29 \\
$\pt \geq 4 \gev$ & 5.37 & 5.30 \\
\bottomrule
\end{tabular}
\end{table}

\begin{table}[t]
\centering
\caption{Average \xf values for the full dataset and the differential fit bins.\label{tab:xf_distr}}
\begin{tabular}{lcc}
\toprule
Bin & $\langle \xf \rangle (\Lc)$ & $\langle \xf \rangle (\Lcbar)$  \\
\midrule
Full range & $-0.128$ & $-0.126$ \\
$\xf < -0.15$ & $-0.270$ & $-0.282$ \\
$-0.15 \leq \xf < -0.06$ & $-0.104$ & $-0.100$ \\
$-0.06\leq \xf<0$ & $-0.038$ & $-0.035$ \\
\bottomrule
\end{tabular}
\end{table}

\begin{figure}[h]
\centering
\includegraphics[width=0.6\textwidth]{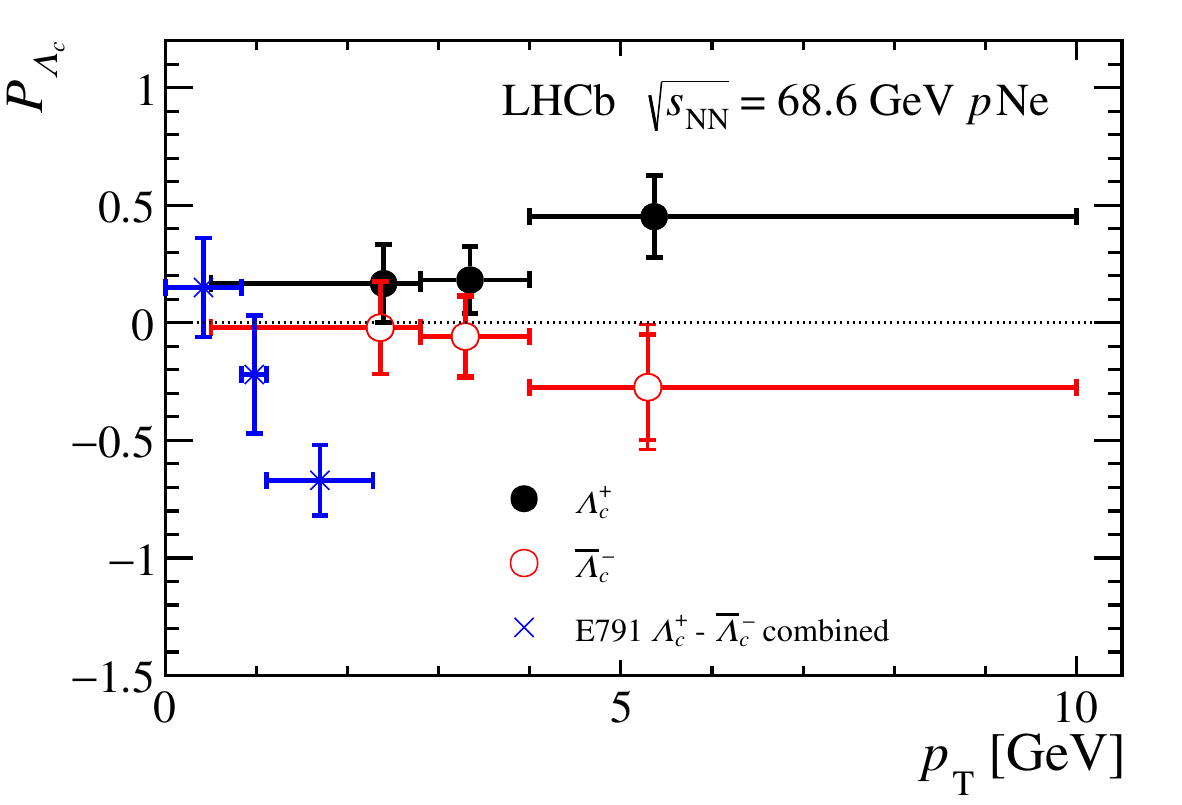}
\caption{Differential polarization measurements in \pt intervals, with (thick line) statistical and (thin line) total uncertainties reported, compared to (blue crosses) E791 experiment results.\label{fig:pol_comp_pt}}
\end{figure}

\begin{figure}[h]
\centering
\includegraphics[width=0.6\textwidth]{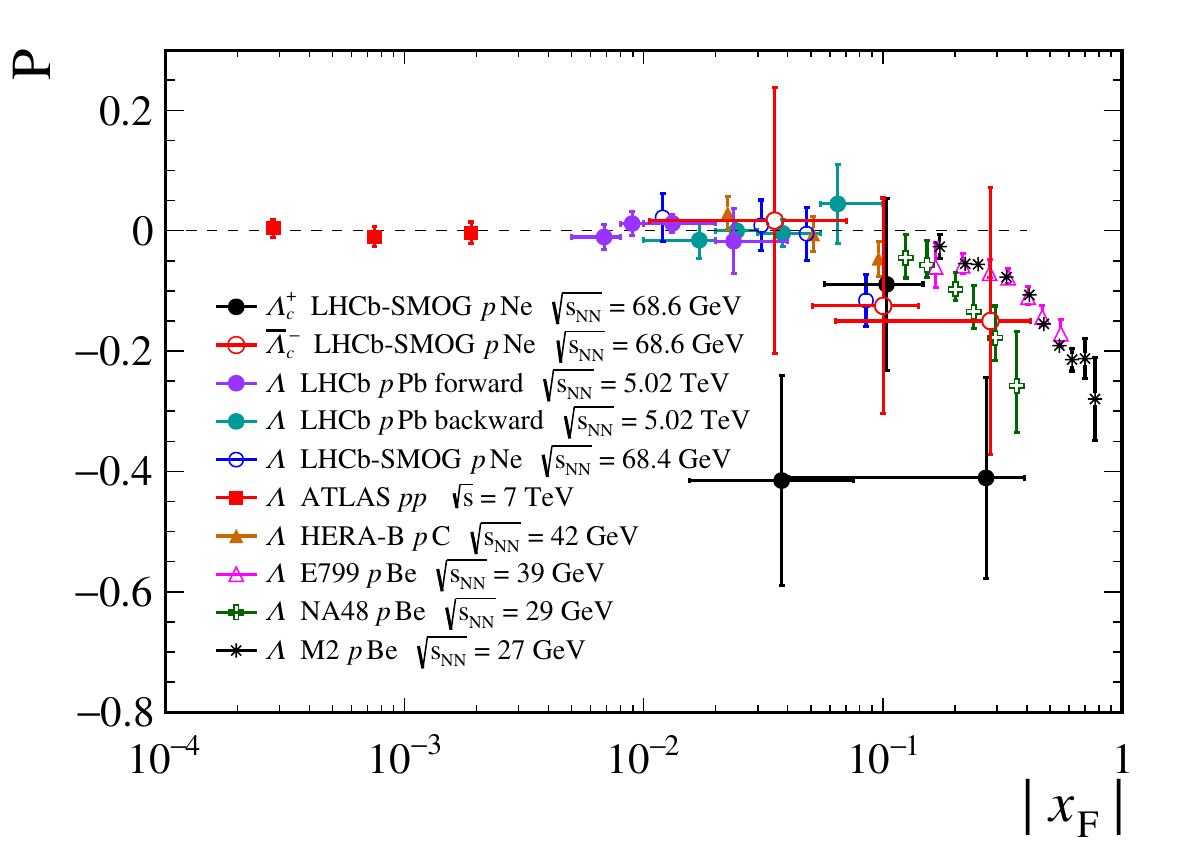}
\caption{Differential polarization measurements in \xf intervals, compared to different \Lz polarization measurements~\cite{Lundberg:1989hw, RAMBERG1994403,na48, Abt_2006, ATLAS:2014ona, LHCb-PAPER-2024-009,LHCb-PAPER-2025-004}. The \Lc polarization results are plotted with an opposite sign for a better comparison, due to the negative \xf coverage of the LHCb SMOG data, and given $P(-\xf) = -P(\xf)$.\label{fig:pol_comp_xf}}
\end{figure}

The differential polarization measurement results as a function of \pt are qualitatively compared to those obtained by the E791 experiment~\cite{Aitala:1999uq} in Fig.~\ref{fig:pol_comp_pt}. Even though both measurements show a possible trend of increasing
polarization with \pt, a quantitative comparison between them is not possible due to the different energy, collision system and polarization measurement technique.
The differential polarization measurement results as a function of \xf are qualitatively compared to \Lz polarization measurements~\cite{Lundberg:1989hw, RAMBERG1994403,na48, Abt_2006, ATLAS:2014ona, LHCb-PAPER-2024-009,LHCb-PAPER-2025-004} in Fig.~\ref{fig:pol_comp_xf}. These results comprise \Lz polarization measurements performed for varying energies and collision systems.

\clearpage

%% file: Authorship_LHCb-PAPER-2025-060.tex
% LHCb collaboration author list
% Data extracted on February 13th, 2026 at 5:13pm for paper reference LHCb-PAPER-2025-060
\centerline
{\large\bf LHCb collaboration}
\begin
{flushleft}
\small
R.~Aaij$^{38}$\lhcborcid{0000-0003-0533-1952},
A.S.W.~Abdelmotteleb$^{58}$\lhcborcid{0000-0001-7905-0542},
C.~Abellan~Beteta$^{52}$\lhcborcid{0009-0009-0869-6798},
F.~Abudin\'en$^{60}$\lhcborcid{0000-0002-6737-3528},
T.~Ackernley$^{62}$\lhcborcid{0000-0002-5951-3498},
A. A. ~Adefisoye$^{70}$\lhcborcid{0000-0003-2448-1550},
B.~Adeva$^{48}$\lhcborcid{0000-0001-9756-3712},
M.~Adinolfi$^{56}$\lhcborcid{0000-0002-1326-1264},
P.~Adlarson$^{86}$\lhcborcid{0000-0001-6280-3851},
C.~Agapopoulou$^{14}$\lhcborcid{0000-0002-2368-0147}
% ...

{\footnotesize \it

% ...
$^{14}$Universit{\'e} Paris-Saclay, CNRS/IN2P3, IJCLab, Orsay, France\\
% ...
$^{38}$Nikhef National Institute for Subatomic Physics, Amsterdam, Netherlands\\
% ...
$^{48}$Instituto Galego de F{\'\i}sica de Altas Enerx{\'\i}as (IGFAE), Universidade de Santiago de Compostela, Santiago de Compostela, Spain\\
% ...
$^{52}$Physik-Institut, Universit{\"a}t Z{\"u}rich, Z{\"u}rich, Switzerland\\
% ...
$^{56}$H.H. Wills Physics Laboratory, University of Bristol, Bristol, United Kingdom\\
% ...
$^{58}$Department of Physics, University of Warwick, Coventry, United Kingdom\\
% ...
$^{60}$School of Physics and Astronomy, University of Edinburgh, Edinburgh, United Kingdom\\
$^{61}$School of Physics and Astronomy, University of Glasgow, Glasgow, United Kingdom\\
$^{62}$Oliver Lodge Laboratory, University of Liverpool, Liverpool, United Kingdom\\
% ...
$^{70}$Syracuse University, Syracuse, NY, United States\\
% ...
$^{86}$Department of Physics and Astronomy, Uppsala University, Uppsala, Sweden, associated to $^{61}$\\
% ...
}
\end{flushleft}